\documentclass{ws-ijmpb}

\begin{document}

\markboth{S.-R. Eric Yang and N.Y. Hwang}
{Edge and bulk merons  in double quantum dots...}

%
\catchline{}{}{}{}{}
%

\title{
Edge and bulk merons  in double quantum dots
with   spontaneous interlayer phase coherence
}

\author{S.-R. Eric Yang and N.Y. Hwang
}
\address{
Physics Department, Korea  University\\
Seoul 136-701, Korea\\
eyang@venus.korea.ac.kr}



\maketitle

\begin{history}
\received{Day Month Year}
\revised{Day Month Year}
\end{history}

\begin{abstract}
We have investigated nucleation of merons
in double quantum dots
when a lateral distortion with a
reflection symmetry is present in the confinement potential.  We find that merons can nucleate both inside and at the edge
of the dots.  In addition to these merons, our results show that electron density modulations can be also present inside the dots.
An edge meron  appears to have approximately a half integer winding number.
\end{abstract}


\section{Model  Hamiltonian}
There  is a range of values of the strength of the confinement potential
for which spontaneous interlayer phase coherence
is  stable in double quantum dots\cite{hu}. Outside this stability range
the shape of the groundstate is expected to undergo a  reconstruction\cite{yang0}.
It is the
competition between the confinement potential and the
Coulomb interaction which drives such a reconstruction.  When the confinement
potential weakens the repulsive Coulomb interaction
pushes electrons from each other, and can create  density depletions inside the
droplet. 

The interplay between the reconstruction and the
formation of merons in the presence of an external potential is
unclear.  
We investigated within a Hartree-Fock (HF) theory how merons
nucleate in the groundstate when a lateral distortion with a
reflection symmetry is present  in the confinement potential\cite{yang0}.  We found that when a sufficiently large distortion
is present merons can be nucleated.
Both density depletions and merons can coexist in the reconstructed states.
In the present work we 
demonstrate by integrating the electron density that these merons have total charge $\pm1/2$.
In addition we 
present  numerical evidence for the presence of edge merons, which
appear to have a half-integer  winding number.

An electron in each dot is in a parabolic potential $\frac{1}{2}m^*\Omega^2r^2$.
In the strong magnetic field limit the energy of an  electron with the angular momentum $m$ in a  quantum dot is
$\epsilon_m=\frac{\hbar\omega_c}{2}+\hbar\omega_p(m+1)$, where
$\omega_c$ is the cyclotron frequency, $\hbar\omega_p=m^*
\Omega^2 \ell^2$, and $\ell$ is the magnetic length.
Hereafter we will measure $\hbar\omega_p$ in units of
$e^2/\varepsilon\ell$, using
$\gamma=\hbar\omega_p/(e^2/\varepsilon\ell)$ which is a dimensionless quantity proportional to
the confinement potential strength  (For GaAs $\gamma=0.131 (\hbar\omega_p[meV])^2/(B[T])^{3/2}$). The
real spins of all electrons are assumed to be frozen aligning with
the magnetic field.
We consider 
a   distortion in the confinement potential of the dot that may be generated by attaching a
metallic stripe vertically on the  cylindrical  metallic gate
surrounding the vertical quantum dots. The vertical distance between two dots is $d$. 
Let us make a simple model for this  type of distortions.
We take a separable form for the distortion potential
$U_{\textrm{dis}}(\vec{r},z)=v g (\vec{r}) h (z)$, where $v$ is the strength of the distortion.
The matrix elements of this potential are
\begin{eqnarray}
\left<n_1 \tau_1|U_{\mathrm{dis}}|n_2 \tau_2 \right> = 
\left(u_{s} \delta_{\tau_1 \tau_2}\!+\! u_{d} (1-\delta_{\tau_1
\tau_2})\right)<n_1|g (\vec{r})|n_2>,
\end{eqnarray}
where  $\left|n_j \tau_j \right>=c_{n_j \tau_j}^{\dagger} \left|0\right>$ denotes an electron with the 
angular momentum $n_j$ in the dot labeled by $\tau_j$.
The parameters $u_{s}$ and $u_{d}$ are defined as 
\begin{eqnarray}
u_{s} &=& v \int \mathrm{d}z h(z) \left|
\varphi\left(z-\frac{d}{2}\right) \right|^2 \; ,\\
u_{d} &=& v \int \mathrm{d}z h(z) \varphi^*\left(z-\frac{d}{2}\right) \varphi \left(z+\frac{d}{2}\right).
\end{eqnarray}
The parameter $u_{s}$ ($u_{d}$) denotes the strength  of the matrix elements of the disorder
potential between intra (inter) layer electrons.
The wavefunctions
$\varphi \left(z+\frac{d}{2}\right)$  and $\varphi\left(z-\frac{d}{2}\right)$ are, respectively,
the left and right dot wavefunctions along the vertical direction.

The total Hamiltonian is $H = H_0 +V$, where
\begin{eqnarray}
H_0 \!&=&\! \sum_{m\sigma} \epsilon_m c_{m \sigma}^\dagger c_{m
\sigma} +\sum_{m\sigma}  \sum_{m'\sigma'}  \left< m \sigma|
U_{\mathrm{dis}}|m' \sigma'\right> c_{m \sigma}^\dagger c_{m'
\sigma'},
\end{eqnarray}
and 
\begin{eqnarray}
V &=& \frac{1}{2}\sum_{m_1 m_2 m_3 m_4} \sum_{\sigma=\uparrow \downarrow} 
\left( \left< m_1 m_2 |V_\mathrm{s}| m_3 m_4 \right>
c_{m_1 \sigma}^\dagger c_{m_2 \sigma}^\dagger c_{m_3 \sigma} c_{m_4 \sigma} \right. \nonumber \\
&& \left. + \left< m_1 m_2 |V_\mathrm {d} | m_3 m_4 \right>
c_{m_1 \sigma}^\dagger c_{m_2,-\sigma}^\dagger c_{m_3,-\sigma}
c_{m_4 \sigma} \right).
\end{eqnarray}
The first term of $H_0$ represents the confining energy due to the
parabolic potential. In the second term, $U_{\mathrm{dis}}$ is the
distortion potential. In the interaction
Hamiltonian $V$, the term $V_s$ ($V_d$) represents the Coulomb interaction
potential between the electrons in the same (different) dots. In
the spatial coordinates, those interactions are given by
$V_s(\vec{r})=e^2/\varepsilon r$ and
$V_d(\vec{r})=e^2/\varepsilon|\vec{r}+d\hat{z}|$, respectively,
where $\varepsilon$ is the dielectric constant.
The spontaneous interlayer phase coherent state (maximum density droplet state\cite{yang1,yang2}),
and meron excitations  are  well approximated by a HF theory\cite{moon}.
The pseudospin density is computed with the calculated  density
matrix.

\section{Results}

\begin{figure}[h]
\begin{center}
\includegraphics[width= 0.5 \textwidth]{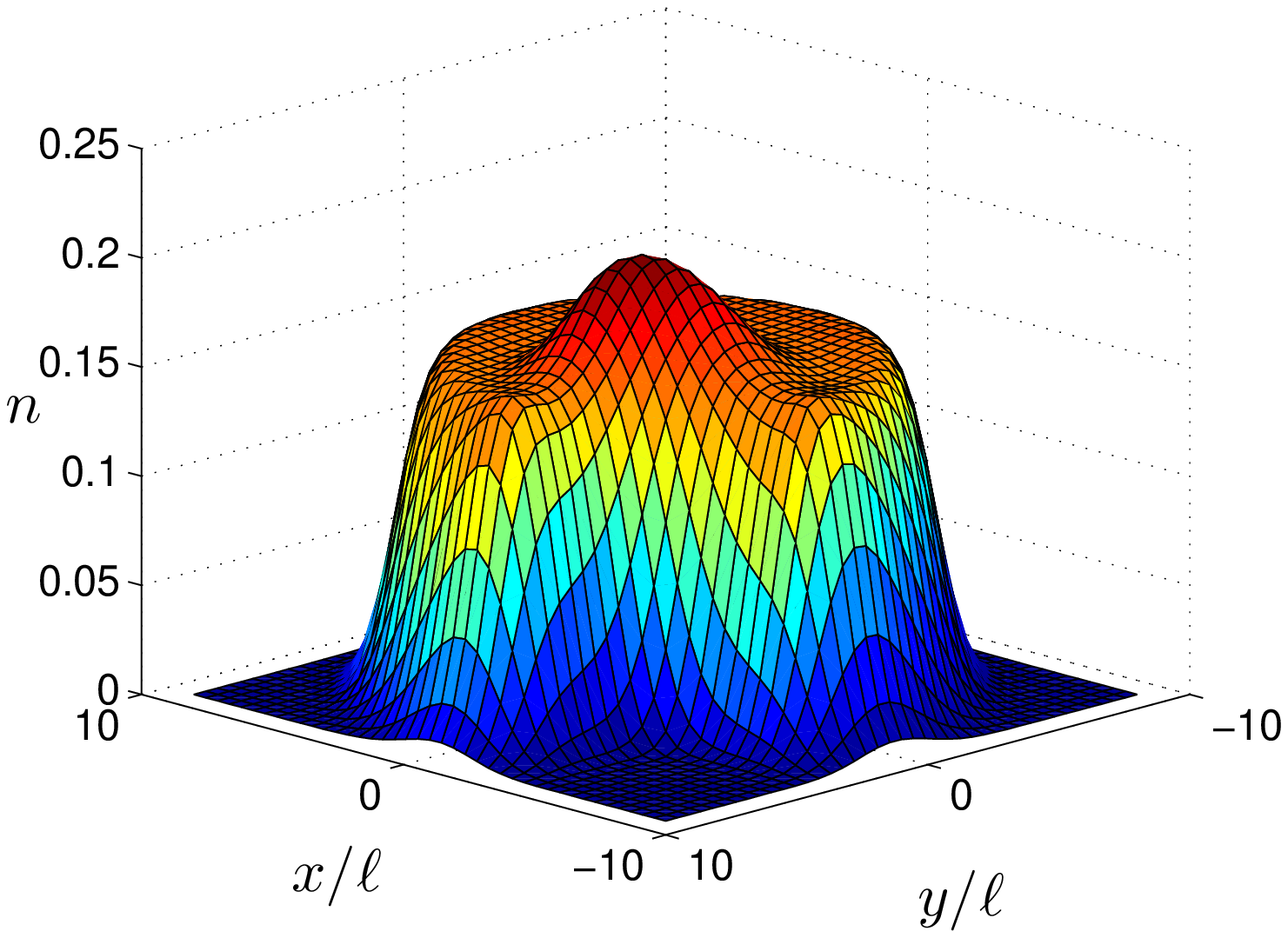}
\includegraphics[width= 0.5 \textwidth]{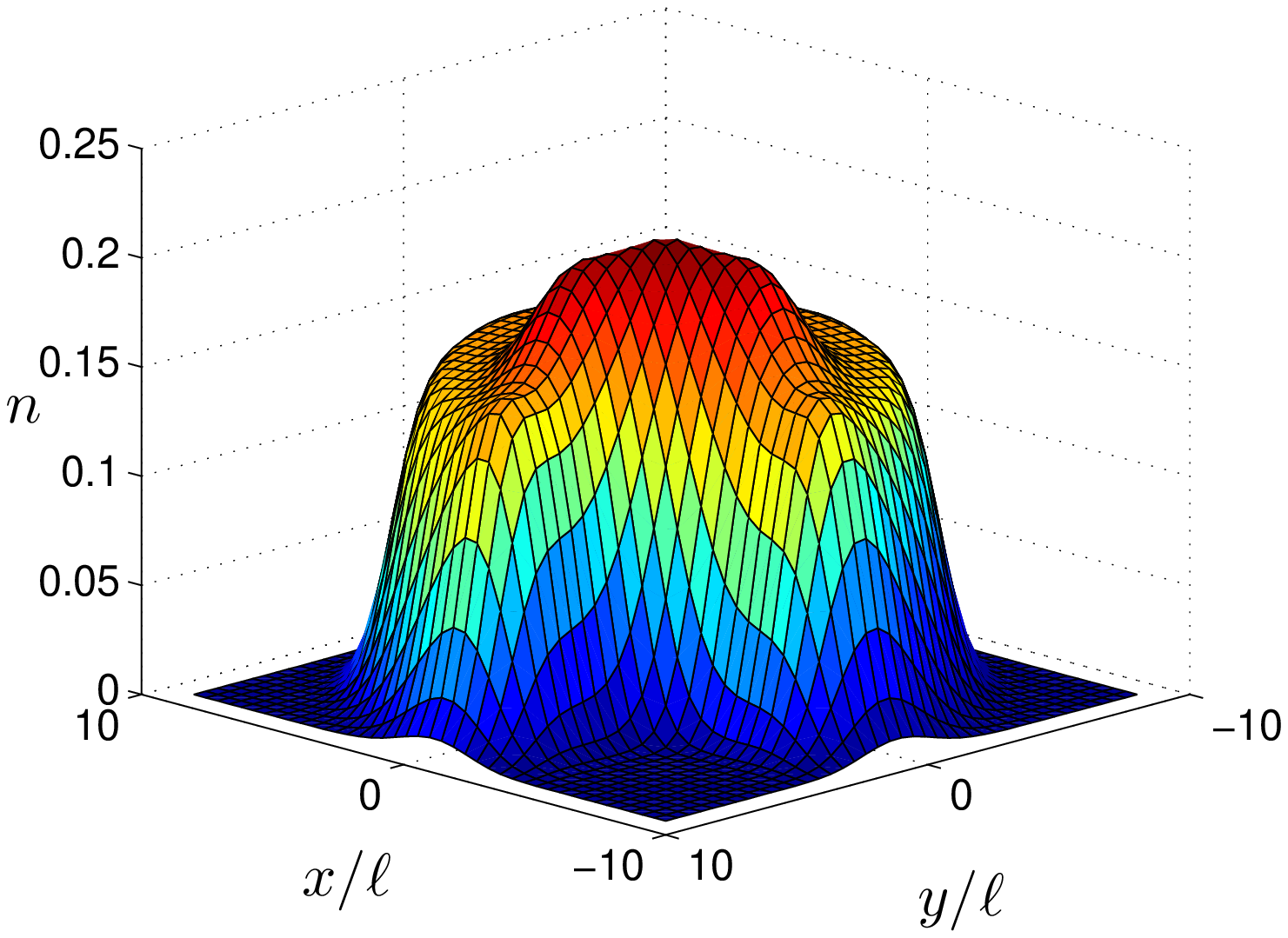}
\caption{Electron densities in units of $1/\ell^2$ for  $(u_{s},u_{d})=(1.5,0.25)(e^2/\varepsilon\ell)$ and
$(u_{s},u_{d})=(1.8,0.3)(e^2/\varepsilon\ell)$.}
\label{fig:density3d.g_0.085.vi_0.50}
\end{center}
\end{figure}

Fig. \ref{fig:density3d.g_0.085.vi_0.50}
 displays total electron densities
for two different strengths of the lateral distortion with the
center at the coordinate $(-5,5)\ell$ by  a Gaussian lateral potential with the range $a=3\ell$.
Results for $\gamma=0.085$ are shown for two different strengths
of the distortion
$(u_{s},u_{d})=(1.5,0.25),(1.8,0.3)(e^2/\varepsilon\ell)$.
The maximum
density is bigger than  $1/2\pi\ell^2$, which implies that
the total local filling factor is bigger than one where the merons are located.  This is consistent
with the presence of merons with negative charges on the background of the total electron
density $1/2\pi\ell^2$.
At  $(u_{s},u_{d})=(1.5,0.25)(e^2/\varepsilon\ell)$ a meron is formed
with the charge $-e/2$ and vorticity $-$.
For the bigger value $(u_{s},u_{d})=(1.8,0.3)(e^2/\varepsilon\ell)$
two merons appear and the
vorticities for the lower and upper merons are $-$ and $+$,
respectively. 
Integrating the electron density  over the range that satisfy $n(r)>\frac{1}{2 \pi\ell^2}$,
we  find the excess charges
to be  
$-0.54 e$ and $-0.98 e$,
which indicates that one meron exists in the upper figure of Fig. \ref{fig:density3d.g_0.085.vi_0.50}
and a meron pair exists in the lower figure of Fig. \ref{fig:density3d.g_0.085.vi_0.50}.
We have attempted to evaluate the total excess charge carried by the merons using the expression\cite{moon} 
\begin{eqnarray}
\Delta N = -\frac{1}{8 \pi} \int {\mathrm d}^2 \bf{r} \
\epsilon_{\mu \nu} \bf{m}(\bf{r}) \cdot [\partial_\mu \bf{m}(\bf{r}) \times \partial_\nu \bf{m}(\bf{r})],
\end{eqnarray}
where the polarization vector $\vec{m}(r)$ is  $\vec{\tau}(r)/n(r)$ ($\vec{\tau}(r)=2\vec{s}/\hbar$ and $\vec{s}$ is the pseudospin).
However, this method gave inaccurate results.   We believe this is because the expression is valid
only for slowly varying pseudospin fields.

\begin{figure}[!htp]
\begin{center}
\includegraphics[width=1.0  \textwidth]{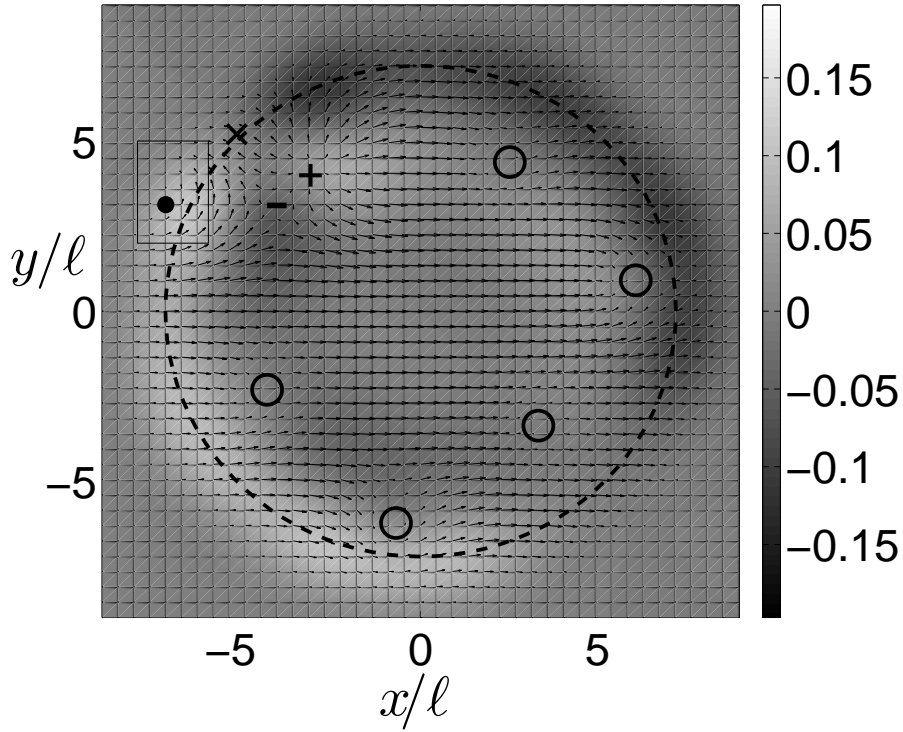}
\caption{ Arrows are projections of  $\vec{\tau}$ on the $xy$ plane
and the gray scale indicates the $z$-component of $\vec{\tau}$.
The dashed circle is the edge of the electron droplet and $+$ and $-$
indicate the core positions of merons with positive and negative winding numbers.
The cross sign is where the impurity is located.
The black dot represents the center of an edge meron, shown in the inset, and
the small circles show electron density depletions.
}
\label{fig:meronpair_one_imp.g_0.066.vi_0.08}
\end{center}
\end{figure}

\begin{figure}[!htp]
\begin{center}
\includegraphics[width=0.6  \textwidth]{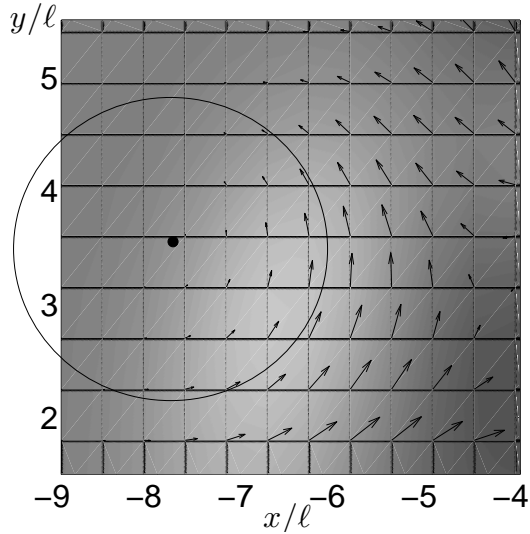}
\caption{
Magnified view of the spin texture of an edge meron with nearly a half-integer winding number.}
\label{fig:lfmeron.one.imp.g_0.066.vi_0.08}
\end{center}
\end{figure}

Fig.~\ref{fig:meronpair_one_imp.g_0.066.vi_0.08} displays how the
pseudospin texture 
for $\gamma=0.066$ and $(u_{s},u_{d})=(0.24, 0.04)(e^2/\varepsilon\ell)$.
We  observe merons inside the dots.
The values of the polarization $m_z(r)=\tau_z(r)/n(r)$ at the meron cores are, within numerical accuracy, $+1$ and $-1$.
This implies that the
charges should be $-e/2$ since the vorticities of these two objects are opposite.  
The total charge of the merons is thus $-e$.
Note that there are also five density depletions inside the dots (the open circles in the figure).  
These  modulations of electron densities are not simply
related to the shape of
$U_{\mathrm{para}}(r)+U_{\mathrm{dis}}(r)$. Since the densities are
inhomogeneous the HF single particle potential must be included into
consideration. It is non-trivial to anticipate the shape of the HF
single particle potential since it reflects the delicate interplay between the
Hartree self energy, exchange self energy, and confinement potential\cite{yang3}.

We also observe  an  edge meron near the edge in Fig.~\ref{fig:meronpair_one_imp.g_0.066.vi_0.08}.
A magnified view of the pseudospin texture is shown in Fig.~\ref{fig:lfmeron.one.imp.g_0.066.vi_0.08}. 
We   draw a circle around the center of this object and see how the in-plane 
pseudospins rotate along the circle.
The in-plane
pseudospins nearly flips as   the half circle is followed inside the dot.
The magnitude of in-plane pseudospin values decreases to zero as the edge region is passed.
We believe thus that the edge meron  has approximately a half integer winding number.
It is complicated to calculate accurately the excess charge of the edge meron since
the electron density varies strongly near the edge even in the absence of an edge meron.
It would be worthwhile to develop an analytical theory for edge merons.

\end{document}